# Direct observation of the influence of the $FeAs_4$ tetrahedron on superconductivity and antiferromagnetic correlations in $Sr_2VO_3FeAs$

Gastón Garbarino[1], Ruben Weht[2,3], Amadou Sow[4], Claudine Lacroix[4], André Sulpice[4], Mohamed Mezouar[1], Xiyu Zhu[5], Fei Han[5], Hai Hu Wen[5], Manuel Núñez-Regueiro[4]

1. European Synchrotron Radiation Facility (ESRF), 6 Rue Jules Horowitz 38043 BP 220 Grenoble
2. Gerencia de Investigación y Aplicaciones, Comisión Nacional de Energía Atómica (CNEA), Avda. General Paz y Constituyentes, 1650 - San Martín, Argentina
3. Instituto Sabato, Universidad Nacional de San Martín - CNEA, 1650 - San Martín, Argentina
4. Institut Néel, Centre National de la Recherche Scientifique (CNRS) & Université Joseph Fourier (UJF), 25 Avenue des Martyrs, F-38042 BP166 Grenoble Cedex 9 France
5. National Laboratory for Superconductivity, Institute of Physics and Beijing National Laboratory for Condensed Matter Physics, Chinese Academy of Sciences, P. O. Box 603, Beijing 100190, China

We measure the pressure dependence of the electrical resistivity and the crystal structure of iron superconductor $Sr_2VO_3FeAs$. Below ~10 GPa the structure compresses but remains undeformed, with regular $FeAs_4$ tetrahedrons, and a constant $T_c$. Beyond 10GPa, the tetrahedron strongly distorts, while $T_c$ goes gradually to zero. Band structure calculations of the undistorted structure show multiple nesting features that hinder the development of an antiferromagnetic ground state (AF), allowing the appearance of superconductivity. The deformation of the tetrahedra that breaks band degeneracy degrades multiple nesting, thus favouring one particular AF state at the expense of $T_c$.

**Short Title**: Tetrahedra, superconductivity and antiferromagnetic correlations in $Sr_2VO_3FeAs$

A frequent trend in the iron pnictides[1] is to have an antiferromagnetic parent compound and a Fermi surface nesting[2], that yields to high temperature superconductivity on doping[3]. On the other hand, as do flat $CuO_2$ planes in cuprates, regular $FeAs_4$ tetrahedra have been empirically shown [4,5,6] to maximize the superconducting transition temperature ($T_c$). Different theoretical models [7,8] based on the strong sensitivity of the bands near the Fermi energy to atomic positions have been advanced to explain this relation. Understanding the differences and similarities between both families of superconductors grants key insight into a general theory of high temperature superconductivity. $Sr_2VO_3FeAs$ [9] has particular characteristics, different from the other Fe pnictides, as the parent stoichiometric compound has a superconducting ground state and shows no evidence of magnetic order in the iron sublattice. Its *FeAs* layers are relatively distant from each other, separated by thick $Sr_2VO_3$ blocks (Fig. 1a). Stoichiometric $Sr_2VO_3FeAs$ with the highest reported $T_c$ has indeed a near regular tetrahedron[4,10,11]. Strong pressure is expected to deform the structure, becoming an excellent probe to understand the anomalous properties of this compound[12], as has been shown in other cases[13].

The polycrystalline samples were synthesized by using a two-step solid state reaction method[9]. Firstly, *SrAs* powders were obtained by the chemical reaction method with *Sr* pieces and As grains. Then they were mixed with $V_2O_5$ (purity 99.9%), *SrO* (purity 99%), *Fe* and *Sr* powders (purity 99.9%), in the formula $Sr_2VO_3FeAs$, ground and pressed into a pellet shape. The weighing, mixing and pressing processes were performed in a glove box with a protective argon atmosphere (the $H_2O$ and $O_2$ contents are both below 0.1 PPM). The pellets were sealed in a silica tube with 0.2 bar of *Ar* gas and followed by a heat treatment at 1150 °C for 40 hours. Then it was cooled down slowly to room temperature.

The electrical resistance measurements were performed using a Keithley 220 source and a Keithley 2182 nanovoltmeter. Pressure measurements, $1.4 - 22 GPa$ (between 4.2K and 300K), were done in a sintered diamond Bridgman anvil apparatus using a pyrophillite gasket and two steatite disks as the pressure medium[14]. Pressure cannot be cycled and thus measurements are done only with increasing pressure.

The angle dispersive X-ray diffraction studies on *Sr$_2$VO$_3$FeAs* powder samples were performed at the ID27 high-pressure beamline of the European Synchrotron Radiation Facility using monochromatic radiation (λ=0.3738Å) and diamond anvil cells. Two transmitting media were used, nitrogen for the room temperature experiment and helium for the low temperature one. The pressure was determined using the shift of the fluorescence line of the ruby. The diffraction patterns were collected with a CCD camera, and the intensity vs. $2\theta$ patterns were obtained using the fit2d software[15]. A complete Rietveld refinement was done with the GSAS-EXPGUI package[16].

The electronic properties for the different structures are analyzed within the Density Functional Theory (DFT) framework. We used the full-potential linearized augmented plane wave code (Wien2k)[17] and GGA[18] to represent the exchange correlation potential. The positions for the "heavy" elements are taken directly from the experiments while oxygen coordinates were optimized. To calculate the weighted real part of the Lindhard susceptibility we evaluate the integral $\chi_Q = \sum_k A_k \cdot A_{k+Q} \cdot (f_k - f_{k+Q})/(\varepsilon_k - \varepsilon_{k+Q} + i\delta)$. $\delta$ is a small energy smearing and $A_k$ and $A_{k+Q}$ the contributions of the *FeAs* layers to the electronic states at ***k*** and ***k+Q***. They are evaluated as the corresponding LAPW projections inside the muffin-tin spheres for the *Fe* and *As* atoms. $\chi_Q$ is evaluated averaging 20 different slices along the $k_z$ axis of the reciprocal space. A fine mesh of *100x100* was used for the $k_x / k_y$ plane.

On Fig. 1b we show the evolution of electrical resistance with pressure on two different samples. They are metallic, following a $T^2$ dependence, with a magnitude that decreases with pressure, while the transition to the superconducting state decreases gradually. Normalizing the resistance with respect to the near ambient value and zooming (Fig. 1c), it is clear that there is a change of the low temperature behaviour starting at pressures around 10GPa, with a less metallic behaviour at high pressures in this region. The detail of the variation of the superconducting $T_c$, presented on Fig. 4a, shows that $T_c$ is almost constant up to ~8GPa decreasing monotonously above, with an extrapolated null value at about 25GPa.

The structural measurements are shown on Fig. 2. Although there seems to be no phase transition to a new crystallographic structure, once more several anomalies are present in the region around 10GPa. There is a kink in the variation of the *c/a* ratio (mid panel of Fig. 2a), that as shown on Fig. 2b corresponds to a sudden halt of the decrease of the *FeAs* distance (upper panel), with almost no variation within error bars of the width of *FeAs* layer (middle panel) or an end to the decrease of the $As_1$-$As_2$ distance (lower panel). All these dependences can be summarized clearer by the variation of the angles of the $FeAs_4$ tetrahedron (Fig. 4b), that pass, within error, from a regular (109°47') and almost constant value below ~10GPa, to irregular values that correspond to a deformed, stretched along the *z* direction, tetrahedron.

A straightforward way to understand the properties of $Sr_2VO_3FeAs$ and the effect of compression is to calculate the electronic properties directly from the actual, measured atomic positions at each particular pressure. Fig. 3a shows the band structures for three different pressures, indicating the contributions of the different subsystems: the states coming from the *FeAs* layers (in red) and from the $Sr_2VO_3$ block (in blue). We will concentrate only on the states coming from the *FeAs* subsystem, supposedly to be the important ones for superconductivity, as suggested by Mazin[19], weighting the electronic states with the corresponding *FeAs* contributions. The respective weighted Fermi surfaces are shown on Fig.

3b. Comparing with other pnictides, the electron FS are no more circular cylinders, but flatter, open vertex square cylinders due to the deformation of the tetrahedra and the partial hybridization with *V* orbitals[14,20,21]. An important parameter to analyse the antiferromagnetic correlations or a spin density wave (SDW) state on these compounds is the real part of the Lindhard electronic susceptibility, $\chi_Q = \sum_k (f_k - f_{k+Q})/(\varepsilon_k - \varepsilon_{k+Q})$, where $f_k$ and $\varepsilon_k$ are the occupation and the energy of the electronic state of wavevector *k*, for a particular wavevector *Q*. It is large when it connects large "parallel" (energy degenerate and thus diverging in $\chi_Q$) portions of the FS. At certain temperature, the electronic system can overreact to any spin wave fluctuation of *Q*, developing a permanent SDW. Doping (addition or subtraction of electrons) modifies the FS, diminishing $\chi_Q$, that below a certain value can no longer trigger the SDW. In the spin fluctuation driven scenario for superconductivity[22], the residual, though still strong, value of $\chi_Q$ is then the main item responsible for a strong pairing interaction generating a high superconducting transition temperature $T_c$.

From our band structures and analysing the weighted $\chi_Q$ it is further clear that the *V* bands and their charge transfer ("self-doping") do not wipe away the characteristic ΓM (*Q*=1/2) peak held responsible for the SDW in pnictides, that remains strong (Fig. 3c). The main outcome of the FS change of shape is that incommensurate nesting peaks in the real part of the bare susceptibility appear at *q'*~1/4, smaller but of comparable magnitude, as shown on Fig. 3c. They are particular to this compound and do not appear in other *Fe* superconductors. Multiple nesting features can be held responsible for the quenching of the itinerant magnetic order, as, plainly speaking, the entropy introduced by the different possible ground states prevents the electron system from developing one of them. Following Moriya's Self Consistent Renormalization theory[23], the effective on site electron repulsion $U_{eff}$ that defines the actual SDW transition temperature $T^*$ when $1/\chi_Q^0(T) < U_{eff}$, becomes $U'_{eff} = U_{eff} - 5gT \sum_{q'} \chi_{q'}$, (g is a positive constant depending on band structure and $\chi_{q'}$ the susceptibilities for the other nestings), effectively lowering the temperature for the appearance of the ordered antiferromagnetic state. The quenching leaves the FS of the compound ungapped down to the temperatures where a superconducting state, originated from the same magnetic interaction, may appear. The peculiar FS of *Sr$_2$VO$_3$FeAs* explains its unusual properties: no magnetic ground state but superconductivity for the undoped material.

Although at this stage it is only a hypothesis, the behaviour of the compound and its band structure at high pressures will give strong support for this scheme as is shown below. Above ~10GPa the *FeAs$_4$* tetrahedra becomes irregular and the $e_g$ derived bands (red in the Fig. 3a) are no more degenerate at the Γ point. The $d_{xy}$ band increases in energy while the $d_{yz}+d_{zy}$ band decreases. The consequence on the FS is that the hole circular cylinders at the Γ point break up, spreading out the incommensurate nesting peaks at ***q'***~1/4, notably diminishing their magnitude. While the commensurate ***Q***=1/2 is broadened, but without losing appreciably its relative weight. To quantify this effect by separating it from the overall variation of the system, the background is subtracted from the amplitude of both peaks. Both $\Delta\chi_{1/2}$ and $\Delta\chi_{1/4}$ are plotted as a function of pressure on Fig. 4c. While $\Delta\chi_{1/2}$ varies slightly, $\Delta\chi_{1/4}$ decreases monotonously, parting away from the former at ~10GPa, which coincides with the beginning of both the deformation of the *FeAs$_4$* tetrahedron and the decrease of $T_c$. As a result the antiferromagnetic ***Q***=1/2 ground state is now favoured, explaining both the low temperature less metallic behaviour above 10GPa, together with the decrease of the superconducting $T_c$. Experimental searching for a low temperature magnetic order at these pressures is presently impossible. We have looked for a possible lattice distortion, as due to fundamental symmetry reasons, the expected SDW order cannot appear in a tetragonal lattice[24]. However, we have not observed any lattice distortion at low temperatures and high pressures, situation that only allows the SDW fluctuations that affect the resistance.

On the other hand, in our analysis the perovskite *V* bands contribute only by seriously deforming the *Fe* bands. This causes an electron charge transfer between *Fe* and *V* bands that we have evaluated by considering the corresponding LAPW projections inside the muffin-tin spheres. The obtained charge transfer is monotonous with pressure and cannot explain the non-monotonous variation of $T_c$. In some recent papers, ferromagnetic ordering of the *V* atoms at low temperatures has been reported[25,26,27,28]. However, this ordering is only observed in samples with low $T_c$ (25K< $T_c$ <35K) at ambient pressure. It has been shown[29] that low $T_c$-s are the result of oxygen deficiencies. These, comparative to ours, low $T_c$ samples also show a remarkably different behaviour at low pressures[26], increasing monotonously up to values similar to ours. The only ARPES measurements existing in the literature [27], shows closed Fermi surfaces different to ours, that support an ordered magnetic state for the *V* atoms, but once more with $T_c$ lower than ours. Also, clear evidence of *V* magnetic ordering [30] has been observed in samples with a $T_c \sim 33K$, that according to Ref. [29] corresponds to a stoichiometry *$Sr_2VO_{2.65}FeAs$*. However, no magnetic ordering has up to date been reported on samples with correct oxygen stoichiometry, $T_c$-s~40K, as is the case of our samples. We claim that stoichiometric samples do not present *V* magnetic ordering, justifying our analysis. Nonetheless, careful measurements with well determined oxygen stoichiometry are necessary to clarify the issue.

Changes of charge transfer under pressure can explain the evolution of superconductivity in a large number of cuprates[31]. Furthermore, measurements of the evolution of $T_c$ under pressure in *$K_xSr_{1-x}Fe_2As_2$* for different *x* show the relevance of this effect in *122* pnictides, as pressure has the same effect as doping as shown in Ref [32]. Thus, it is necessary to check the importance of this effect in *$Sr_2VO_3FeAs$*. However, the calculated evolution of charges as a function of pressure is monotonous, and cannot explain a $T_c$ pressure dependence with the abrupt change observed at ~10$GP$a, although it could be important in a quantitative calculation of $T_c$.

To verify if the deformation of the tetrahedra is the reason behind the $\Delta\chi_{1/4}$ change, the electronic band structure corresponding to a putative crystal with the 19.47GPa *a* and *c* lattice parameters but with an undeformed tetrahedron was calculated. The calculated $\Delta\chi_{1/4}$ falls now on the same curve as the $\Delta\chi_{1/2}$, showing that the decrease is effectively due to the deformation of the tetrahedron.

If the *q'*~1/4 incommensurate peaks quenches the SDW state switching the antiferromagnetic fluctuations strength towards superconductivity, it should be possible to scale the dependence of $T_c$ with pressure on that of $\Delta\chi_{1/4}$ as a parameter. One possible phenomenological scaling ($V \sim \Delta\chi_{1/4}$) is shown on Fig. 4d. The correlation between the two trends is clear, taking into consideration that, from our calculations, the *Fe* projected $N(E_F)$ is almost constant as a function of pressure. Furthermore, the value calculated for the undeformed tetrahedron simulation falls at a value near to the ambient pressure $T_c$.

In conclusion, we have measured the evolution with pressure of superconductivity and crystallographic structure, calculating the electronic properties for the actual positions at each pressure. Contrary to other pnictides, we find here multiple nesting that should play an important role in antiferromagnetism and superconductivity. Above ~10 GPa, we observe that the regular *FeAs₄* tetrahedron strongly distorts, breaking band degeneracy and degrading the multiple nesting of the Fermi surface, with an simultaneous effect on $T_c$, that goes gradually to zero. Our data provide a transparent explanation for the dependence of $T_c$ with the regularity of the *FeAs₄* tetrahedron for this material, while strongly supporting the importance of the antiferromagnetic correlations for superconductivity.

Acknowledgements

This work was partially supported by the project TetraFer ANR-09-BLAN-0211 of the Agence Nationale de la Recherche of France. R.W. is member of CONICET-Argentina and gratefully acknowledges partial support from CONICET (Grant No. PIP 112-200801-00047) and ANPCyT (Grant No. PICT 837/07).

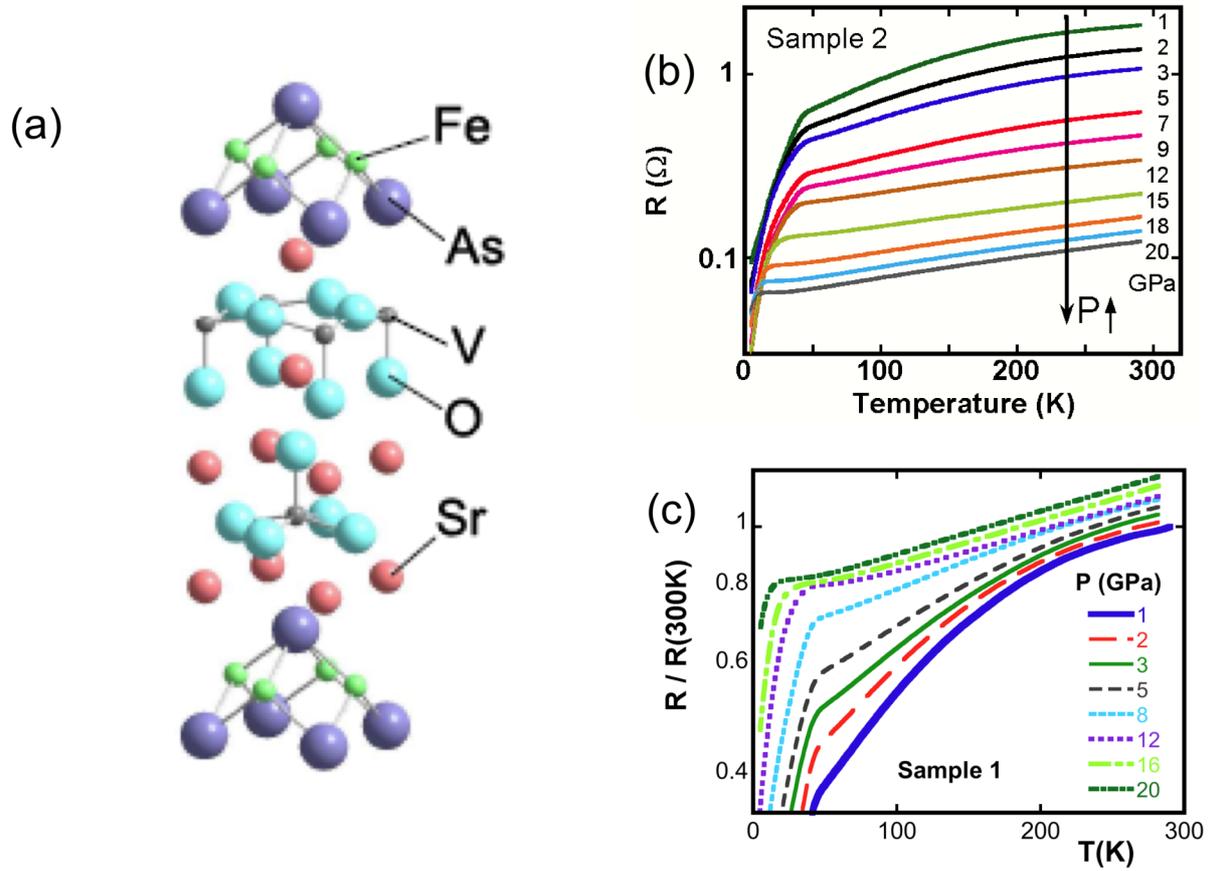

**Figure 1**

(a) The crystal structure of the *Sr₂VO₃FeAs* compound, showing the perovskite thick layer sandwiched between *FeAs* layers. (b) Electrical resistance of sample 2 as a function of temperature with pressure as a parameter. The decrease in temperature of the superconducting transition is visible. (c) Electrical resistance in logarithmic scale of sample 1 normalized to its room temperature value (curves are uniformly shifted for clarity). The passage, at low temperatures, with increasing pressure from a metallic correlated electron $T^2$ behaviour to a less metallic behaviour, presumably due to scattering against SDW fluctuations, is clear.

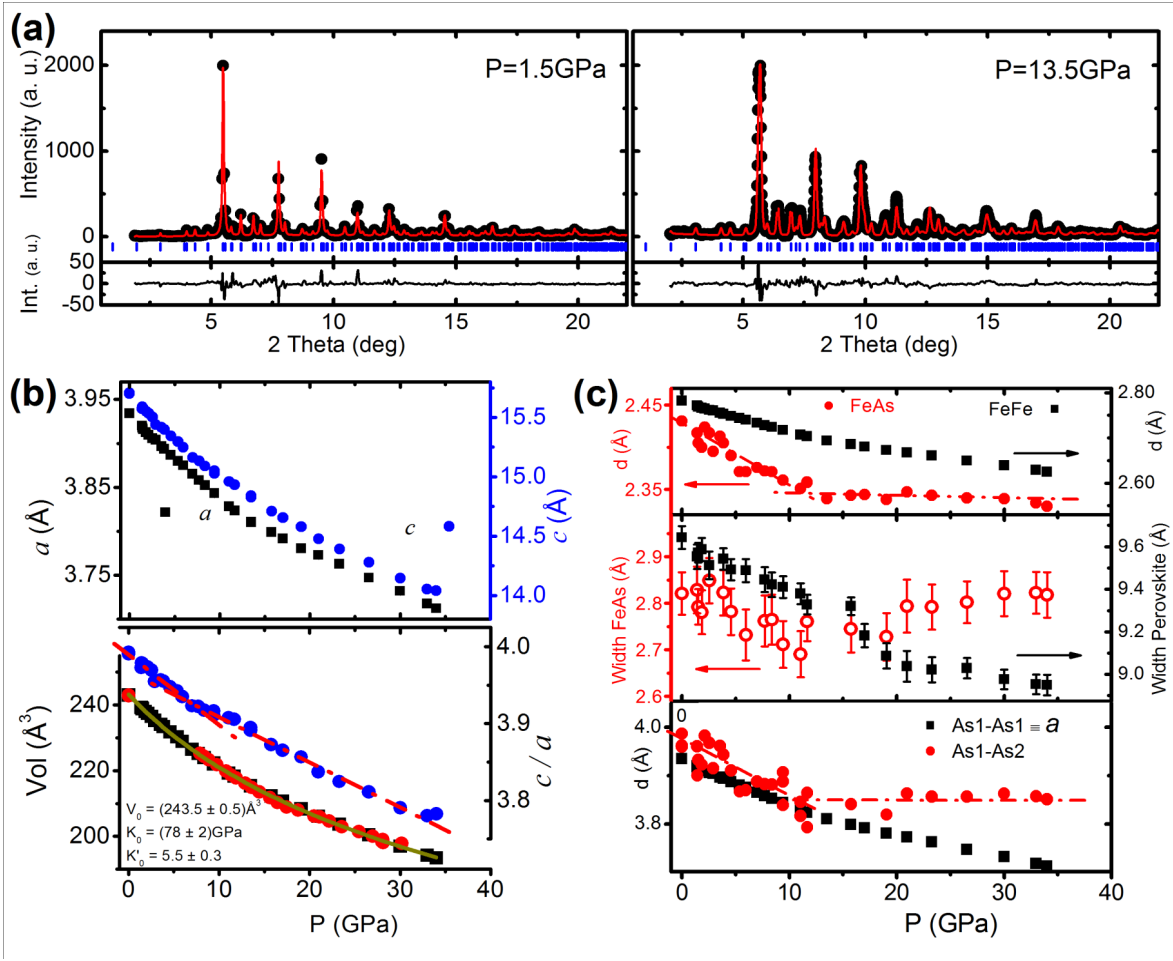

Figure 2 (a) X-ray diffraction pattern measured and calculated (black dots and red lines, respectively) at $1.5 GPa$ and $13.5 GPa$. The blue lines represent the reflections of the tetragonal $P4/nmm$ structure. (b) Upper panel: evolution with pressure of the lattice parameters **a** (black squares) and **c** (blue circles); middle panel: evolution of the **a**/**c** ratio, the kink due to a deformation in compression is visible; lower panel: evolution with pressure of the volume showing the values of an third order Birch Murnaghan equation of state, where $K_0$ is the bulk modulus and $K'_0 = \partial K_0/\partial p$ its pressure dependence. (c) Evolution of several characteristic distances as a function of pressure, showing the parameters responsible for the kink at ~10GPa.

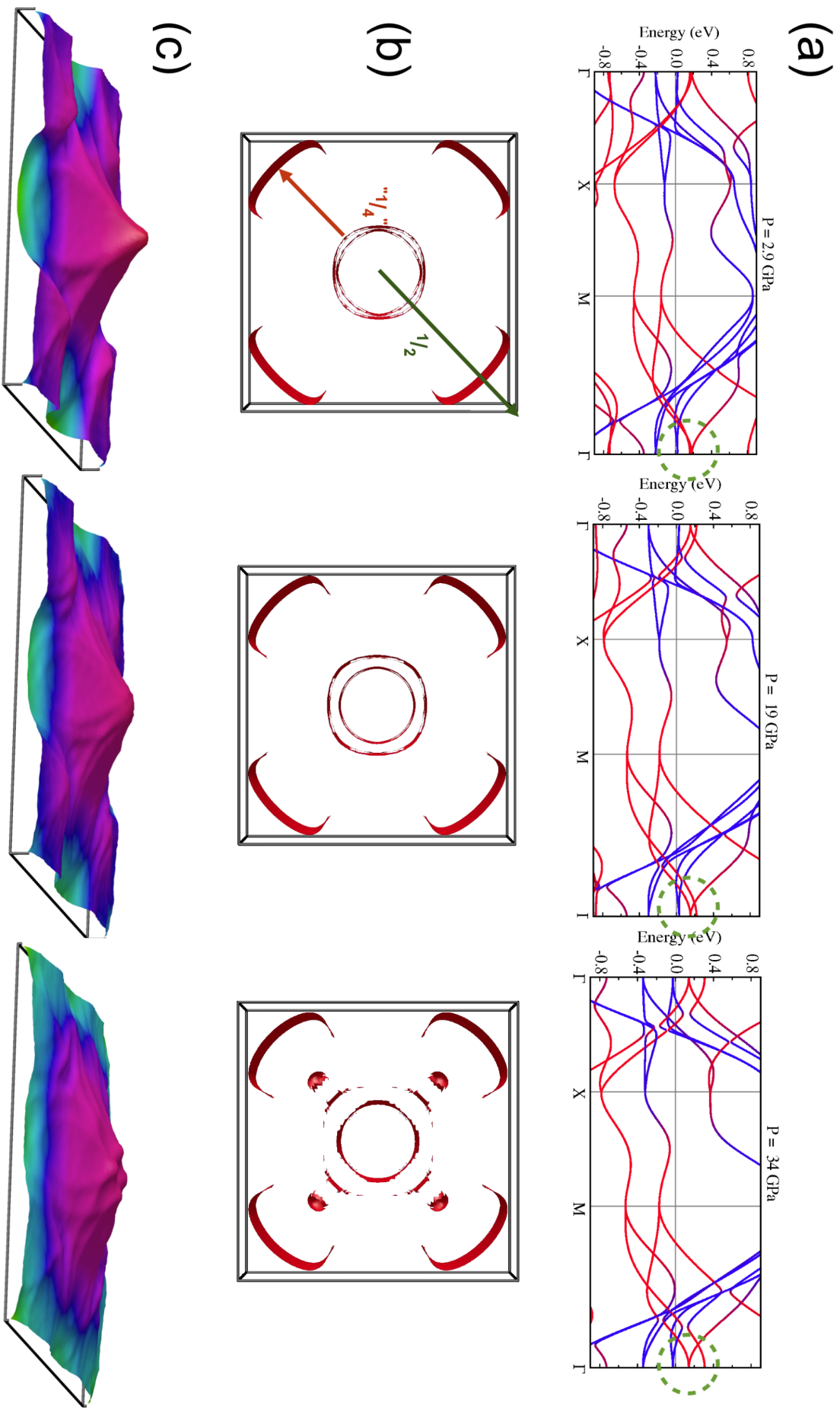

**Figure 3**

(a) Electronic band structures at three different pressures. *Fe* (*V*) orbital derived bands are in red (blue), the dashed green circles emphasize the splitting of the $d_{xy}$ and ($d_{yz}+d_{zy}$) bands at $\Gamma$ due to the deformed *FeAs₄* tetrahedra. (b) *Fe* orbital FS at the same three pressures (the total *Fe+V* are shown on Fig. 3S), showing the two nesting wavevectors, ***Q***=1/2 and ***q'***~1/4. (c) Real part of the weighted bare electronic Lindhard susceptibility $\chi_q$ in the Brillouin zone (BZ), showing the central ***Q***=1/2 peak together with the four ***q'***~1/4 peaks that decrease with pressure.

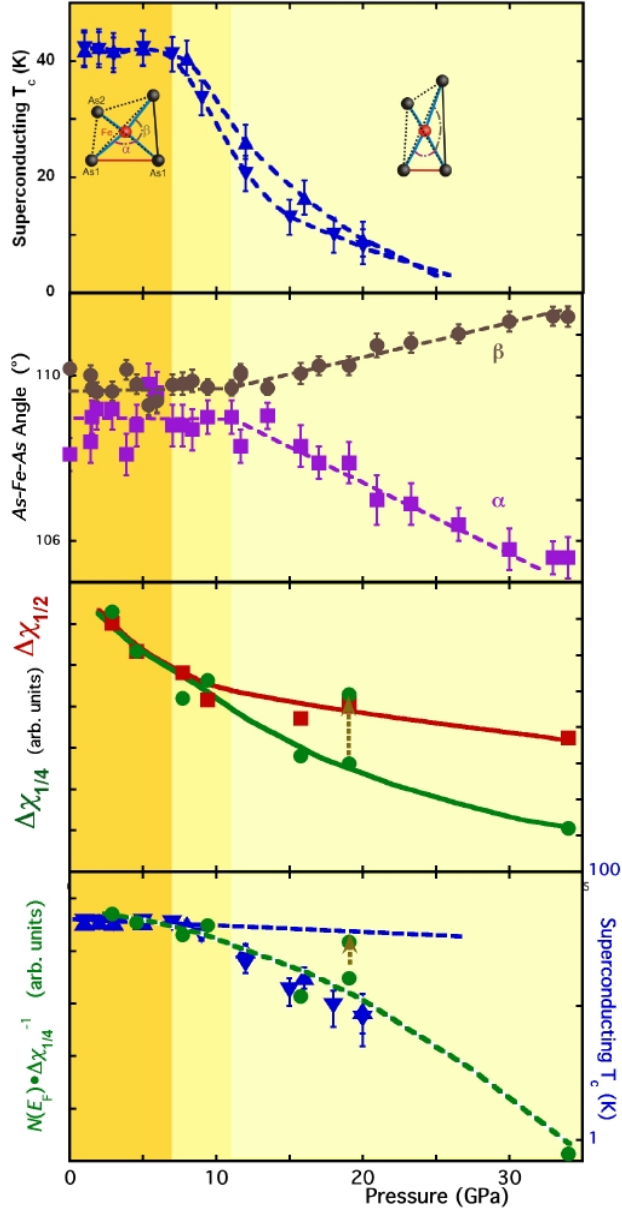

**Figure 4**

(a) Dependence on pressure of the superconducting onset $T_c$ for two samples. (b) Evolution with pressure of the two principal angles ($\alpha$ and $\beta$ as defined on the drawing), showing that up to ~10GPa the tetrahedron is regular. (c) Change with pressure in the relative magnitude ($\Delta\chi_q = \chi_q - bckg$) for the two principal nesting wavevectors. The arrow shows how $\Delta\chi_{1/4}$ changes in a putative undeformed tetrahedron structure. (d) Scaling of $T_c$ on $\Delta\chi_{1/4}$ using the phenomenological relation $\ln T_c \sim \left[ N(E_F) \Delta\chi_{1/4} \right]^{-1}$.

[1] Y. Kamihara, T. Watanabe, M. Hirano. & H. Hosono J. Am. Chem. Soc. 130, 32963297 (2008).
[2] F. Ma and Z-Y. Lu, *Phys. Rev.* B 78, 033111 (2008)
[3] I.I. Mazin and J. Schmalian, Physica C 469, 614-627(2009)
[4] C-H.Lee et al., *J. Phys. Soc. of Japan* **77**, 083704 (2008)
[5] A. Kreyssig et al., Phys. Rev. B78, 184517(2008)
[6] H. Okabe, N. Takeshita, K. Horigane, T. Murakana and J. Akimitsu, *Phys. Rev.* B81, 205119(2010)
[7] K. Kuroki, H. Usui, S. Onari, R. Arita and H. Aoki , *Phys. Rev.* B **79**, 224511 (2009).
[8] M.J. Calderón, B. Valenzuela and E. Bascones, *New J. Phys.* **11,** 013051 (2009).
[9] X. Zhu et al., 105001 (2008)
[10] X. Zhu et al.,, *Phys. Rev.* B79, 220512(R), 2009.
[11] F. Han et al., *Sci China Ser* G, 53 1202 (2010)
[12] H. Kotegawa et al., *J. Phys. Soc. Japan* 78,123707(2009)
[13] S.A.J. Kimber et al., , *Nature Mat.* 8,471 (2009)
[14] S. Sanfilippo etal., *Phys. Rev. B*, **61** R3800(1998).
[15] A.P. Hammersley, S.O. Svensson, M. Hanfland, A.N. Fitch and D. Hausermann, *High Pressure Res.*, **14** 235-248 (1996)
[16] A.C. Larson and Von Freele, *Los Alamos National Laboratory Report* LUAR p 86-748 (1994); B.H.Toby, *J.Appl.Cryst.*, **34** 210 (2001)
[17] P. Blaha, K. Schwarz, G.K.H. Madsen, D. Kvasnicka and J. Luitz, in *WIEN2K, An Augmented Plane Wave and Local Orbitals Program for Calculating Crystal Properties*, edited by K. Schwarz (Vienna, University of Technology, Austria, (2001)
[18] J.P. Perdew, K. Burke and M. Ernzerhof, *Phys. Rev. Lett.* 77, 3865-3869 (1996)
[19] I.I. Mazin, *Phys. Rev.* B81, 020507(R), 2010.
[20] K-W. Lee and W.E. Pickett, *Europhys. Lett.* 89, 57008 (2010)
[21] G. Wang, M. Zhang, L. Zheng and Z. Yang, *Phys. Rev.* B 80, 184501(2009)
[22] T. Moriya and K. Ueda, *Rep. Prog. Phys.* **66**, 1299–1341 (2003)
[23] T. Moriya "*Spin Fluctuations in Itinerant Electron Magnetism*", Springer Verlag (Berlin), (1985)
[24] A. Cano, M. Civelli, I. Eremin and I. Paul, *Phys. Rev.* B 82, 020408(R) (2010)
[25] J. Munevar et al., ArXiv Cond-mat 1011.1894), 2010
[26] H. Kotegawa et al., ArXiv Cond-mat 1009.5491), 2010
[27] T. Qian et al. , *Phys. Rev. B*, **83** R140513 (2011)
[28] S. Tatematsu, E. Satomi, Y. Kobayashi and M. Sato, *J. Phys. Soc. Japan* **79**,123712(2010)
[29] F. Han et al., Sci China Ser G, 53, 1202 (2010)
[30] M. Tegel et al. , *Phys. Rev. B*, **82** R1405à7 (2010)
[31] For a review see : M. Núñez-Regueiro and C. Acha, *High pressure measurements on mercury cuprates, in Studies of high temperarue superconductors*, Edit. A. Narlikar, Nova Science Publishers (New York) Vol. 24, 203 (1997)
[32] M. Gooch, B. Lv, B. Lorenz, A.M. Guloy and C-W. Chu, *Phys. Rev.* B78, 180508(2008)